# A Perturbation Theory Modification of the Flory-Huggins Polymer Solution Theory

by

**G. Ali Mansoori**
University of Illinois at Chicago
(M/C 063) Chicago, IL 60607-7000 USA
Email: <mansoori@uic.edu>

## Abstract

A perturbation theory modification of the Flory-Huggins polymer solution theory is presented. The proposed perturbation equation utilizes the results by Tukur *et al* [J. Chem. Phys. 110 (7), 3463, 1999] for hard-sphere binary mixture at infinite size ratio. The resulting perturbation theory equations are used to predict properties of three different polymers with different molecular weights in different solvents. Comparison of the proposed perturbation calculations with those of the Flory-Huggins theory and the experimental data indicate that the proposed perturbation method appreciably improves prediction of polymer solution properties especially at large polymer / solvent size ratios.

# Introduction

Various statistical mechanical theories of polymer solutions have been developed during the last half a century. The original and one of the best known of these theories is the Flory-Huggins theory [1,2]. It is shown that the Flory-Huggins theory is a rather crude approximation for polymer solutions when the polymer to solvent size ratio increases [3]. There have been several attempts to improve the predictive capability of the Flory-Huggins theory [4-9] by empirical modification of its interaction parameter $(\chi)$. In the perturbation modification of the Flory-Huggins theory presented here the original expression of the polymer solution theory and its interaction parameter is retained. However, additional terms are added to the Flory-Huggins equation correcting its asymptotic value at infinite size ratio and at infinite temperature.

According to the Flory-Huggins theory [10] for the mixture of a polymer and a solvent the following expression holds for the Gibbs free energy of mixing $\Delta G_M$:

$$\Delta G_M^{FH} = RT(n_1 \ln \varphi_1 + n_2 \ln \varphi_2 + \chi_1 n_1 \varphi_2), \tag{1}$$

where, $n_i$ and $\varphi_i$ are number of moles and volume fraction of component $i$, respectively, and $\chi_1$ is the Flory-Huggins interaction parameter. According to Eq. (1), the expressions for activities of solvent (1) and polymer (2) are as the following:

$$\ln a_1^{FH} = \ln \varphi_1 + \left(1 - \frac{1}{r}\right)\varphi_2 + \chi_1 \varphi_2^2, \tag{2}$$

$$\ln a_2^{FH} = \ln \varphi_2 - (r-1)(1-\varphi_2) + \chi_1 r(1-\varphi_2)^2. \tag{3}$$

In the above equations $a_i^{FH}$ and $\varphi_i$ are activity and volume fraction of component $i$, *respectively,* and $r$ is segment fraction which is the ratio of molecular volume of polymer to molecular volume of solvent i.e. :

$$r \equiv (v_2/v_1) = (\sigma_2/\sigma_1)^3, \tag{4}$$

where, $v_1$, $v_2$ are molecular volume and $\sigma_1$ and $\sigma_2$ are molecular diameters of solvent and polymer, respectively. The Flory-Huggins interaction parameter $(\chi_1)$ has the following definition:

$$\chi_1 \equiv z\Delta w_{12} x_1 / kT, \tag{5}$$

which is a dimensionless quantity [10] that characterizes the interaction energy per solvent molecule divided by $kT$. In Eq. (5) $z$ is coordination number and $\Delta w_{12}$ is interchange energy.

From the expressions of the activities, Eqs. (2) and (3), the following equations for the excess Gibbs free energy over an ideal solution, entropy of dilution and enthalpy of dilution of the polymer and solvent for the binary polymer plus solvent mixture are derived :

$$\frac{G^E_{FH}}{RT} = x_1 \left[ ln\,\varphi_1 + \left(1 - \frac{1}{r}\right)\varphi_2 + \chi_1 \varphi_2^2 - ln\,x_1 \right] + x_2 \left[ ln\,\varphi_2 - (r-1)\varphi_1 + \chi_1 r \varphi_1^2 - ln\,x_2 \right] \quad (6)$$

According to Eqs. (2), (3) and (5) :

$$\Delta \overline{S}_1^{FH} = -R \left[ ln\,\varphi_1 + \left(1 - \frac{1}{r}\right)\varphi_2 \right] \quad (7)$$

$$\Delta \overline{S}_2^{FH} = -R \left[ ln\,\varphi_2 - (r-1)\varphi_1 \right] \quad (8)$$

and also:

$$\Delta \overline{H}_1^{FH} = RT \varphi_2^2 \chi_1 \quad (9)$$

$$\Delta \overline{H}_2^{FH} = RTr \varphi_1^2 \chi_1 \quad (10)$$

In the above equations $\Delta \overline{S}_i^{FH}$ and $\Delta \overline{H}_i^{FH}$ are entropy of dilution and enthalpy of dilution of component $i$.

Considering the Eq. (1) it can be shown that:

$$\lim_{\sigma_i \to 0} \left( \frac{\partial G^E_{FH}/RT}{\partial \sigma_i} \right)_{T,P,\sigma_{j \neq i}} = 0 \quad (11)$$

This equation indicates that according to the Flory-Huggins theory the partial derivative of the excess Gibbs free energy of the mixture at the hard sphere limit ( $T \to \infty$ ) and at infinite size ratio limit ( $\sigma_i \to 0$ ) diminishes

Recently Tukur et al [11] showed that for a binary hard sphere mixture of infinite size difference the following expression is rigorously valid for partial derivative of the excess Helmholtz free energy, $A^E$, of the polymer+solvent mixture with respect to the molecular diameter of solvent when the solvent diameter approaches to zero:

$$\underset{\sigma_i \to o}{limit} \left( \frac{\partial A_{hs}^E / RT}{\partial \sigma_i} \right)_{T,V,\sigma_{j \neq i}} = \frac{\frac{\pi}{2} N_A \rho x_i x_j \sigma_j^2}{1 - \frac{\pi}{6} N_A \rho x_j \sigma_j^3} \qquad (12)$$

where $\rho$, $N_A$, $x_i$, $\sigma_i$ and $T$ are mixture density, Avogadro's number, mole fraction of component $i$, molecular size of component $i$ and absolute temperature, respectively. This expression can be applied for the excess Gibbs free energy, $G^E$, assuming the excess volume of the mixture is negligible, i.e.

$$\underset{\sigma_i \to o}{limit} \left( \frac{\partial G_{hs}^E / RT}{\partial \sigma_i} \right)_{T,P,\sigma_{j \neq i}} = \frac{\frac{\pi}{2} N_A \rho x_i x_j \sigma_j^2}{1 - \frac{\pi}{6} N_A \rho x_j \sigma_j^3} \qquad \text{for} \qquad v^E = 0. \qquad (13)$$

In the following section we utilize Eq. (13) in a perturbation correction of the Flory-Huggins theory of polymer solutions.

## Perturbation Modification of the Flory-Huggins Theory

Considering that the infinite temperature limit is practically equivalent to the hard-sphere limit and comparing Eqs. (11) and (13) it is obvious that the hard-sphere limit (T→∞) at the infinite size ratio ($\sigma_1$→0) as predicted by the Flory-Huggins equation is not correct. This indicates that the Flory-Huggins expression for excess Gibbs is the first term of the double perturbation expansion of excess Gibbs with respect to powers of ($\sigma_1$) and (1/T),

$$\frac{G^E}{RT} = \frac{G_{FH}^E}{RT} + \left( \frac{\partial \left( \frac{G_{FH}^E}{RT} \right)}{\partial \sigma_1} \right)_{\sigma_1 = 0} \cdot \sigma_1 + \left( \frac{\partial \left( \frac{G_{FH}^E}{RT} \right)}{\partial \left( \frac{1}{T} \right)} \right)_{T = \infty} \cdot \left( \frac{1}{T} \right) + higher\ order\ terms \qquad (14)$$

Now, by integrating Eq. (13) with respect to $\sigma_1$ and (1/T) and replacing for the second and third terms in Eq. (14) and neglecting the higher order terms, of the orders of $O[\sigma_1]^2$, $O[\sigma_1 \cdot (1/T)]$, $O[1/T]^2$ and higher, the following expression for the perturbation modification of the Flory-Huggins excess Gibbs free energy will be derived [12,13]:

$$\frac{G^E}{RT} = \frac{G_{FH}^E}{RT} + \left\{ 3 \left[ (\eta_1 / (1 - \eta_1))(s - 1) / x_1 + (\eta_2 / (1 - \eta_2))(s^{-1} - 1) / x_2 \right] \right. \qquad (15)$$

$$\left. + \left[ \alpha \cdot (\eta_1 (E \cdot s - 1) / x_1 + \eta_2 (s^{-1} - 1) / x_2) \right] \frac{1}{RT} \right\} x_1 x_2$$

In this equation

$$\alpha = f \varepsilon_2, \quad E = \varepsilon_2 / \varepsilon_1. \qquad (16)$$

$s$ represents the ratio of molecular diameters of polymer to solvent, i.e.

$$s \equiv \sigma_2 / \sigma_1 = r^{1/3} \equiv (v_2 / v_1)^{1/3}, \tag{17}$$

and $\eta_i$ is the packing fraction of component $i$ in the mixture,

$$\eta_i \equiv \frac{\pi}{6} N_A \rho . x_i \sigma_i^3, \tag{18}$$

Eq. (15) can predict the exact asymptotic value given by Eq. (13). In what follows Eq. (15) is used to calculate properties of various polymer solutions with different polymer/solvent size ratios and comparisons are made with the Flory-Huggins theory. It is shown that the proposed perturbation expression is capable of predicting polymer solution properties more accurately than the Flory-Huggins theory when the size difference between the polymer and solvent increases. Starting with Eq. (15), expressions for the other thermodynamic properties of polymer solution can be derived:

The expressions for the activity of solvent, $a_1 = x_1 \gamma_1$, and activity of polymer, $a_2 = x_2 \gamma_2$, are derived as the following [12]:

$$\ln a_1 = \ln a_1^{FH} + \left\{ 3\left( \frac{\eta_1 / x_1 (s-1)}{(1-\eta_1)^2} + \frac{\eta_2 / x_2 (s^{-1} - 1)(1 - \eta_2 / x_2)}{(1-\eta_2)^2} \right) + \right. \tag{19}$$

$$\left. \frac{\alpha}{RT} \left( \eta_1 / x_1 (Es - 1) + (s^{-1} - 1)\eta_2 / x_2 \right) \right\} x_2^2$$

and

$$\ln a_2 = \ln a_1^{FH} + \left\{ 3\left( \frac{\eta_1 / x_1 (1 - \eta_1 / x_1)(s-1)}{(1-\eta_1)^2} + \frac{\eta_2 / x_2 (s^{-1} - 1)}{(1-\eta_2)^2} \right) + \right. \tag{20}$$

$$\left. \frac{\alpha}{RT} \left( \eta_1 (Es - 1) / x_1 + \eta_2 (s^{-1} - 1) / x_2 \right) \right\} x_1^2$$

since $\ln \gamma_i = \left( \partial \left( nG^E / RT \right) / \partial n_i \right)_{T,P,n_{j \neq i}}$ and $a_i = x_i \gamma_i$. In these equations $a_1^{FH}$ and $a_2^{FH}$ are the Flory-Huggins solvent and polymer activities, respectively as given by Eqs. (2) and (3). Expressions for activities of solvent and polymer and Eq. (5), can be used to derive the following equation for the entropy of dilution, $\Delta S_i = -R[\partial(T \ln a_i)/\partial T]_{P,\varphi_2}$ of solvent due to addition of the polymer:

$$\Delta \bar{S}_1 = \Delta \bar{S}_1^{FH} - 3R\left(\frac{\eta_1/x_1(s-1)}{(1-\eta_1)^2} + \frac{\eta_2/x_2(1/s-1)(1-\eta_2/x_2)}{(1-\eta_2)^2}\right)x_2^2 \qquad (21)$$

and

$$\Delta \bar{S}_2 = \Delta \bar{S}_2^{FH} - 3R\left(\frac{\eta_1/x_1(1-\eta_1/x_1)(s-1)}{(1-\eta_1)^2} + \frac{\eta_2/x_2(1/s-1)}{(1-\eta_2)^2}\right)x_2^2 \qquad (22)$$

Also from the expressions for the activities the following equations for the heat of dilutions, $\Delta \bar{H}_i = -RT^2(\partial \ln a_i/\partial T)_{P,\varphi_2}$, are derived:

$$\Delta \bar{H}_1 = \Delta \bar{H}_1^{FH} + \alpha[(1/s-1)\eta_2/x_2 + (Es-1)\eta_1/x_1]x_2^2 \qquad (23)$$

and

$$\Delta \bar{H}_2 = \Delta \bar{H}_2^{FH} + \alpha[(1/s-1)\eta_2/x_2 + (Es-1)\eta_1/x_1]x_1^2 \qquad (24)$$

In the following section, the above equations are used to calculate properties of various polymer solutions and the results are compared with the calculations based on the Flory-Huggins relations and experimental data.

## Calculations and discussions

In order to test the present perturbation theory model three different polymer+solvent systems with various (solvent/polymer) size ratios, for which experimental data are available [14-16], are used. These systems (Benzene+PDMS3,850, Toluene+PS290,000 and MEK+PS290,000) are identical to those used in the original paper of Flory [10]. In the first stage parameters $E$ and $\sigma_1$ of the polymers are calculated [12] using a group contribution method [17,18]. Parameters $\chi_1$, $\sigma_2$ and $\alpha$ are determined [12] using experimental solvent activity data using Eq. (19). Similarly parameters $\chi_1$ and $\sigma_2$ of the Flory-Huggins theory are calculated from the experimental activity data. Numerical values of all these parameters for the three polymer-solvent mixtures are reported in Table 1.

The enthalpy of dilution of the solvents in polymer solution systems are calculated using Eqs. (9) and (23). The result of calculations are shown in Figures 1-3. According to these figures the proposed perturbation modification generally improves the predictions. As the size difference between polymer and solvent increases improved predictions by the perturbation model are more pronounced.

The entropy of dilution of solvents in polymers of the same three systems are predicted and the results are reported in Figure 4 based on the present perturbation model and the Flory-Huggins theory. According to this figure, the entropy of dilutions predicted by the Flory-Huggins equation are not in agreement with the experimental data specially for high molecular weight polymers while the calculations by the perturbation model are in very good agreement with the experimental data for, both, low and high molecular weight polymers. The entropy of dilutions predicted by the Flory-Huggins theory, for high molecular weight polymer solutions (Toluene+PS290,000 and MEK+PS290,000) are in error, nearly constant, and independent of the polymer under consideration.

The results reported in this paper demonstrate the major improvement in predicting polymer solution properties using the proposed perturbation modification of the Flory-Huggins theory. The improvement becomes more appreciable for polymer solutions with large size difference between the polymer and solvent.

**Acknowledgements:** This research is supported in part by MSTRI. The author would like to thank Dr. A. Eliassi, Prof. E,Z. Hamad, Prof. H. Modarress and Mr. N.M. Tukur for their helpful communications.

**Table 1. Parameters used for the present study**

| system | group-cont. method | | Flory-Huggins | | Perturbation model | | |
|---|---|---|---|---|---|---|---|
| | $\sigma_1$ [nm] | $E$ [-] | $\sigma_2$ [nm] | $\chi_1$ [-] | $\sigma_2$ [nm] | $\chi_1$ [-] | $\alpha$ [ J / mol ] $\times 10^{-3}$ |
| Benzene + PDMS 3,850 | 0.535 | 1.620 | 1.895 | 0.809 | 1.500 | 0.426 | -36.271 |
| Toluene + PS 290,000 | 0.574 | 1.602 | 7.797 | 0.343 | 6.300 | 0.114 | 806.077 |
| MEK + PS 290,000 | 0.539 | 1.480 | 7.732 | 0.725 | 6.800 | 0.149 | 693.142 |

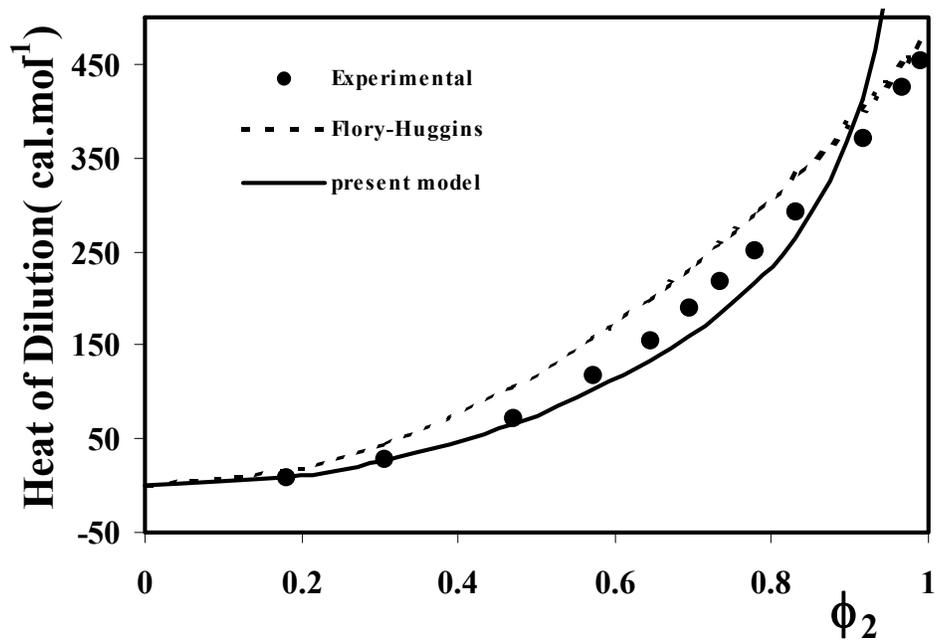

**Figure 1:** Solvent (1) heat of dilution versus volume fraction of polymer (2) for the solution of benzene (solvent) in PDMS3,850 (polymer) at 25 °C. The solid circles are the experimental data [19], the dashed lines are the calculations based on the Flory-Huggins theory and the solid lines are the calculations based on the present perturbation model.

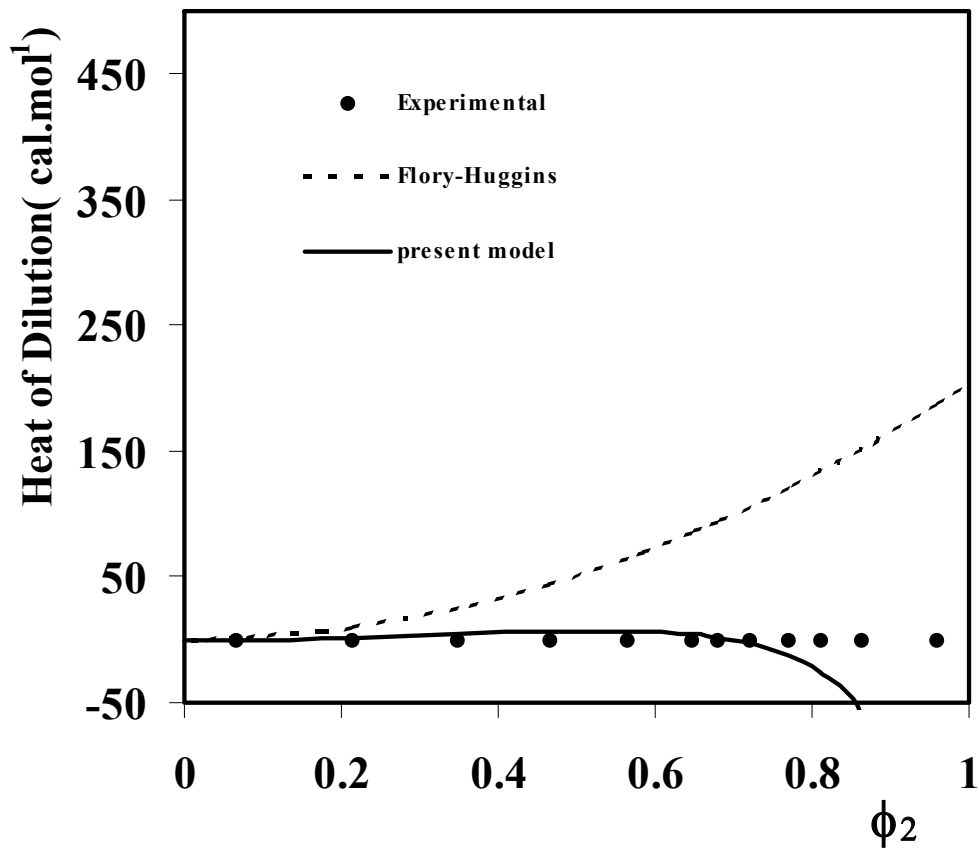

**Figure 2:** The same as Figure (1) but for the system toluene (solvent) in PS290,000 (polymer) and the experimental data is taken from Ref. [20].

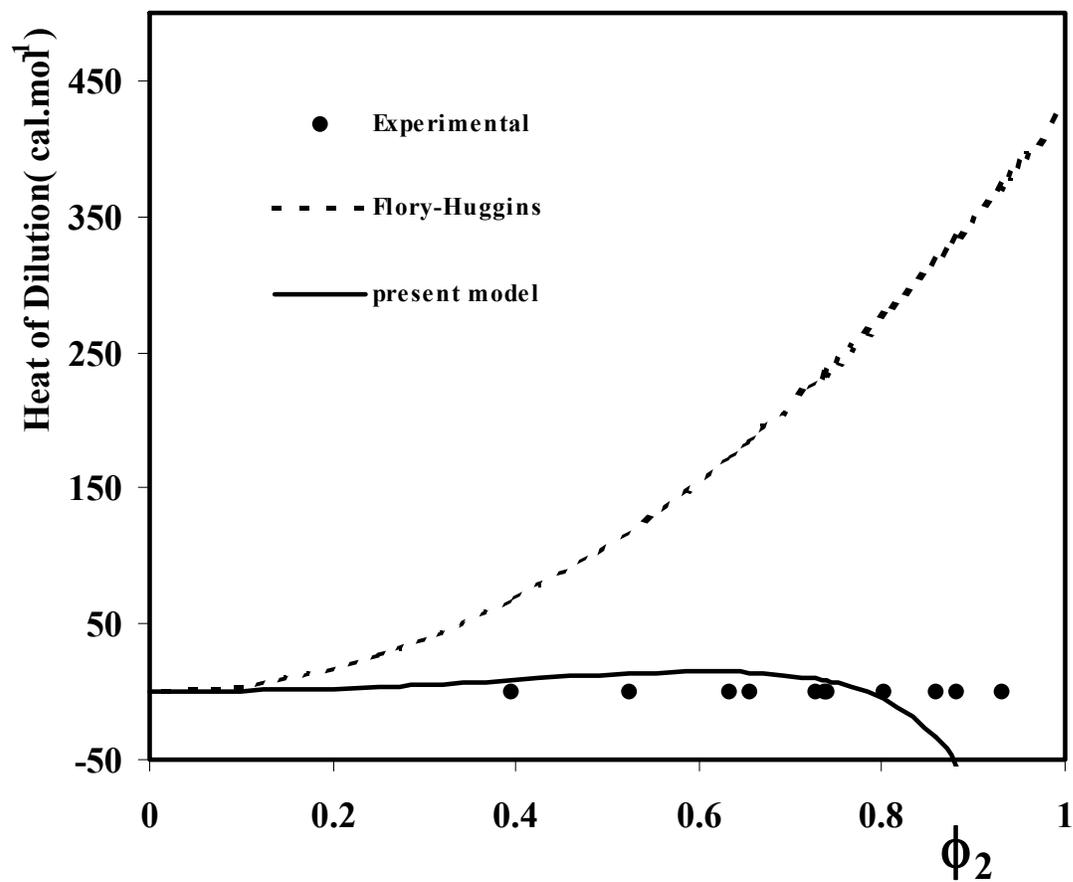

**Figure 3:** The same as Figure (1) but for the system MEK (solvent) in PS290,000 (polymer) and the experimental data is taken from Ref. [20].

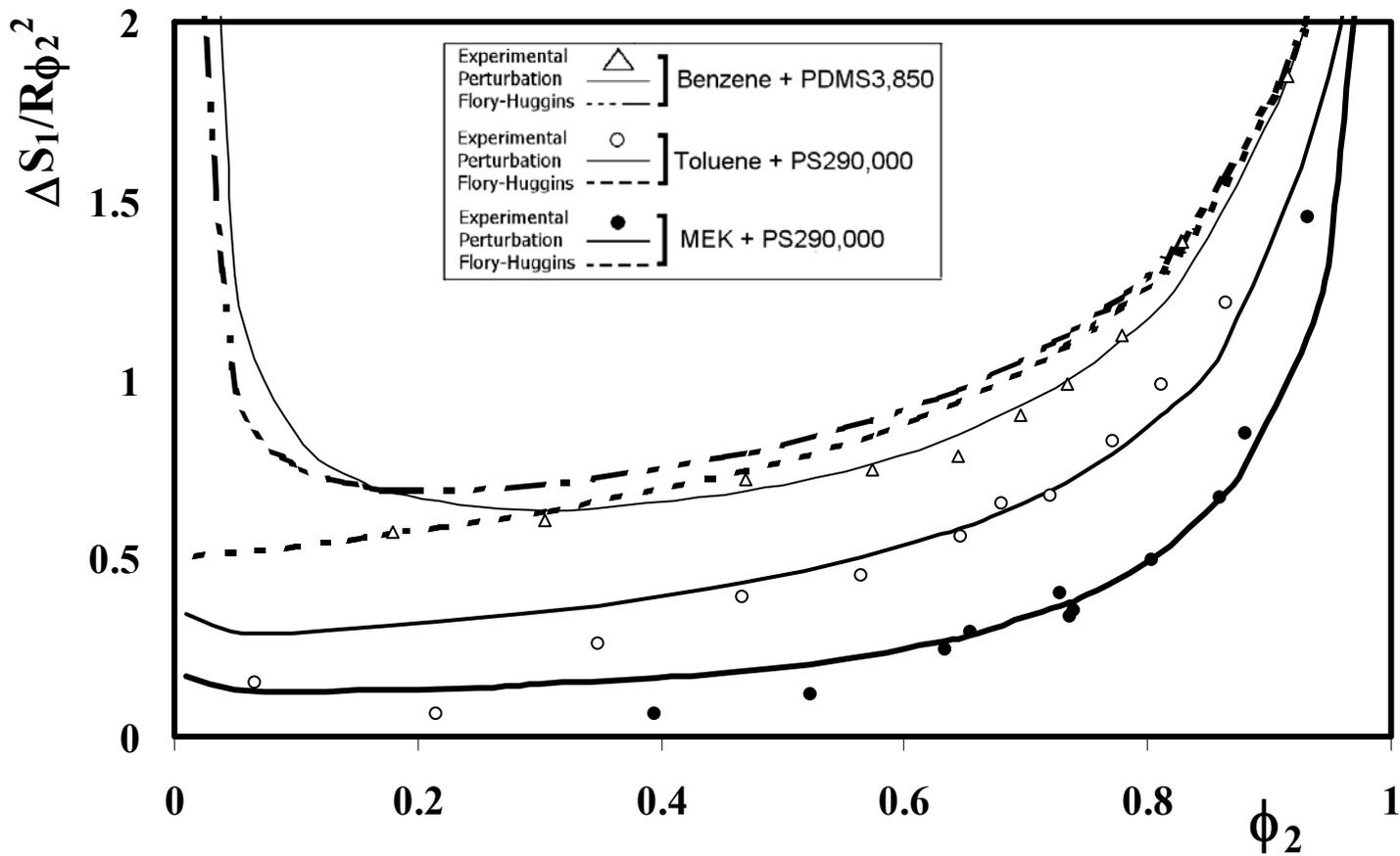

**Figure 4:** Solvent (1) entropy of dilution versus volume fraction of polymer (2) for three different solvent+polymer mixtures (Benzene+PDMS3,850, Toluene+PS290,000 and MEK+PS290,000) at 25 $^{o}$C . In this figure predictions by the present perturbation model are compare with the Flory-Huggins theory and the experimental data [19,20].